# Coulomb contribution to Shockley-Read-Hall recombination

Konrad Sakowski[1,2*], Pawel Strak[2], Pawel Kempisty[2], Jacek Piechota[2], Izabella Grzegory[2], Piotr Perlin[2], Eva Monroy[3], Agata Kaminska[2,4,5], Stanislaw Krukowski[2]

[1]*Institute of Applied Mathematics and Mechanics, University of Warsaw, Banacha 2, 02-097 Warsaw, Poland*

[2]*Institute of High Pressure Physics, Polish Academy of Sciences, Sokolowska 29/37, 01-142 Warsaw, Poland*

[3]*Univ. Grenoble-Alpes, CEA, Grenoble INP, IRIG, PHELIQS , 17 av. des Martyrs, 38000 Grenoble, France*

[4]*Institute of Physics, Polish Academy of Sciences, Aleja Lotnikow 32/46, PL-02668 Warsaw, Poland*

[5]*Cardinal Stefan Wyszynski University, Faculty of Mathematics and Natural Sciences. School of Exact Sciences, Dewajtis 5, 01-815 Warsaw, Poland*

Abstract

Defect-mediated nonradiative recombination, known as Shockley-Read-Hall (SRH) recombination is reformulated. The introduced model considers Coulomb attraction between charged deep defect and the approaching free carrier, showing that this effect may cause considerable increase of the carrier velocity approaching the recombination center. The effect considerably increases the carrier capture rates. It is demonstrated that, in a typical semiconductor device or semiconductor medium, SRH recombination cannot be neglected at low temperatures. We also show that SRH is more effective in the case of low doped semiconductors. Effective screening by mobile carrier density could reduce the effect leading to SRH rate increase.

*Corresponding author, email: konrad@unipress.waw.pl

I. INTRODUCTION

Shockley-Read-Hall (SRH) recombination is a non-radiative process which involves a defect and an electron-hole pair. SRH is one of the most important detrimental processes for optoelectronic devices, having a negative impact on the performance of all semiconductor light emitters, including laser diodes (LDs), light emitting diodes (LEDs), superluminescent light emitting diodes (SLEDs), and electroluminescent lamps. [1, 2] Therefore, investigations of the basic properties of SRH recombination are of great interest from the fundamental point of view and also for applications. The basic features of the process were described relatively early, independently by Shockley and Read, [1] and by Hall. [3] Since then the primary role of SRH process in the description of the energy dissipation and emission processes remains unchanged. [4, 5, 6]

The recent advent of nitride optoelectronic confirmed the important role of SRH recombination. However, the optical properties of these materials present a higher degree of complexity due to the initial high density of defects, alloy inhomogeneity, polarization, complex Auger processes and strong electron-phonon interaction. Hence, the peculiarities of the SRH process were difficult to assess, and the role of SRH recombination in the overall performance is still under discussion. [7, 8, 9, 10] It is widely recognized that, in InGaN LEDs, carrier localization in In-rich areas prevents migration to dislocation lines, which usually serve as effective nonradiative SRH recombination centers in semiconductors. [7, 8, 9] Such a positive insulation effect is not observed in AlGaN-based devices, which leads to a lower emission efficiency in comparison to their InGaN-based counterparts. [11]

The role of SRH recombination is universally recognized despite its relatively simple model. It is assumed that a deep defect level, i.e. the state located deep into the bandgap, is occupied by the majority carrier. As the energy of the state is much below the conduction band minimum of the crystal host and also much below Fermi energy level, we can assume that the occupation of the deep localized state is close to complete. In the SRH process, the minority carrier is "captured" by the defect to recombine. This is the slowest, rate controlling step, therefore SRH is effectively a mono-molecular process with a recombination rate that is proportional to the density of minority carriers. The net result is the thermalization of the excess energy in the form of lattice vibrations, i.e. phonons that effectively heat the sample. Thus, the effect is doubly detrimental, first by the loss of the energy and second by the increase of temperature, which decreases device efficiency, and can ultimately lead to degradation of the device.

The SRH recombination model, devised by Shockley, is based on a classical description of the minority carrier collision with the deep defect occupied by the majority carrier. [1, 3, 4] The recombination rate is calculated from the average capture rate of the minority carrier by the defect. This collision cross-section depends on the defect type (point or linear). Regardless of the type of defect, a shared feature of both types is that the size of the deep defect falls within the order of a lattice constant $a$. The cross-section is calibrated to experimental data, sometimes supplemented by the capture probability factor. This description has facilitated a basic modeling of optoelectronic devices albeit with reduced prediction accuracy.

SRH recombination rates can be extracted from time-resolved photoluminescence (TRPL) measurements. However, directly determining the SRH rate from TRPL data is challenging because the effect is entangled with the other processes, including radiative or Auger recombination. [12] Recently, we proposed a new method of analysis of TRPL data which allows disentangling mono-, bi- and tri-molecular contributions to the overall photoluminescence decay. [13, 14] However, the obtained results do not support simple attribution of mono-molecular processes to SRH recombination, bi-molecular to radiative and tri-molecular to Auger processes. [13] Further investigation of the nature of these processes and their contribution to overall decay is required. In the present work, we study the carrier capture processes showing that Coulomb attraction can considerably increase SRH rate.

## II. RESULTS

Deep defects have their eigenstates localized in a length scale of a few Angstroms. These eigenstates serve as conversion centers from electron-hole pairs to phonons. In the SRH recombination event, the deep center is assumed to be occupied by the majority carrier. Then, the SRH process is reduced to minority carrier capture that leads to its annihilation with the localized majority carrier. Here, we delve into this capture process.

Following Shockley, we consider first the standard SRH formulation, where it is assumed that minority carriers and holes travel over a length $l$ before being captured by the deep defect, with average constant thermal velocity:

$$v_{th} = \sqrt{\frac{kT}{m_{eff}}} = \sqrt{A} \tag{1}$$

where $m_{eff}$ is the minority carrier effective mass. The newly introduced parameter $A = v_{th}^2 = \frac{kT}{m_{eff}}$ will be used later. The carrier capture cross-section S is estimated assuming that the size of the deep defect is of the order of the lattice constant $a$. [14] Thus, the carrier capture cross-section is: $S_{def} = \pi a^2$ and $S_{dis} = a$ for the capture by the point defect and by the dislocation line, respectively. The capture condition is that the carrier encounters a single defect at the capture length $l$, i.e. $lN_{def}S_{def} = 1$ or $lN_{dis}S_{dis} = 1$, where point defect and dislocation density are denoted by $N_{def}$ and $N_{dis}$, respectively. Note that in general, $lS_{def}$ ($lS_{dis}$) corresponds roughly to volume (area, respectively) swept by the carrier travelled over distance $l$ in presence of point defects (dislocation lines), and thus $lN_{def}S_{def}$ ($lN_{dis}S_{dis}$) is the average number of encountered defects (dislocations). Therefore, the benchmark capture lengths $l$ associated with the point defects are:

i/ for $N_{def} = 10^{17}\ cm^{-3}$  $\qquad l = 3.13 \times 10^{-5}\ m$,

ii/ for $N_{def} = 10^{19}\ cm^{-3}$ $\qquad l = 3.13 \times 10^{-7}\ m$,

The benchmark capture lengths $l$ associated with the dislocations are:

iii/ for $N_{dis} = 10^{7}\ cm^{-2}$ $\qquad l = 3.13 \times 10^{-2}\ m$,

iv/ for $N_{dis} = 10^{10}\ cm^{-2}$ $\qquad l = 3.13 \times 10^{-5}\ m$,

For other defect densities these values scale linearly, thus it may be easily assessed. It is worth to note that these values are relatively large, much bigger than the size of quantum structures such as quantum wells. They are rather of the order of the typical cavity length of laser diodes. Note also that the lengths are lower for point defects that indicate on dominant role of the latter in SRH recombination.

The combination of the capture length $l$ and the thermal velocity $v_{th}$ provides a standard estimate of the average SRH capture time:

$$\tau_{SRH}(T) = \frac{l}{v_{th}} = l\sqrt{\frac{m_{eff}}{kT}} \qquad (2)$$

Thus, the variation of the SRH rate from low temperature (5 K) to room temperature is $\tau_{SRH}(5K)/\tau_{SRH}(300K) \approx 7.75$, i.e. less than one order of magnitude for the decrease from room temperature to liquid-helium temperature. This is not a drastic decrease, nevertheless it is frequently assumed that the SRH can be neglected for the temperatures close to that of liquid helium. [15, 16, 17, 18]

The above comparison of the capture time at low and room temperature was based on the assumption that carriers travel with constant thermal velocity in the crystal. However, a considerable fraction of the deep defects is charged by captured majority carriers, thus the approaching minority carrier is attracted by a Coulomb force. As such, it is accelerated towards the defect. Additionally, the charged defect can be partially screened by the free carrier gas, which reduces the effect. Free-carrier screening is characterized by the screening length $\lambda$.

As mentioned above, the carrier capture cross-section is relatively small, of the order of the lattice parameter $a$. Thus, only carriers that move directly towards the defect are captured, and such process can be modelled using a one-dimensional approximation. In order to estimate the capture time, we use the energy conservation law, according to which the sum of kinetic and potential energies is conserved:

$$\frac{m_{eff}v^2}{2} + e\,V(r) = constant \tag{3}$$

where $V(r) = -\frac{e\,exp(-r/\lambda)}{4\pi\varepsilon_o\varepsilon\,r}$ is the Coulomb screened potential for a carrier charge equal to the elementary charge, $e$. As the mass of the defect is at least three orders of magnitude larger than the mass of free carriers, the kinetic energy of the defect can be neglected. The velocity change on the path from infinity to a point at the distance $r$ from the charged defect can be calculated as:

$$v(r) = \sqrt{\frac{kT}{m_{eff}} + \frac{e^2 exp(-r/\lambda)}{2\pi\varepsilon_o\varepsilon\,r m_{eff}}} \tag{4}$$

where it was assumed that the carrier velocity at infinity is equal to the thermal velocity, i.e. $v(\infty) = v_{th} = \sqrt{\frac{kT}{m_{eff}}}$. We also assume that the distance from the center $r$ in (4) is not very close to $r = 0$, as then Newtonian physics approximation is no longer valid for this system. The carrier velocity along the path is therefore position dependent:

$$v(r) = \sqrt{A + B\left[\frac{exp(-r/\lambda)}{r}\right]} \tag{5}$$

The parameters in the above equation are: $A = \frac{k_BT}{m_{eff}} = v_{th}^2$ and $B = \frac{e^2}{2\pi\varepsilon\varepsilon_o m_{eff}}$. The latter is material dependent only, and was calculated here for GaN ($\varepsilon_{GaN} = 10.28$). [19, 20, 21, 22] In the case of the electron capture ($m_{eff-GaN} = 0.2\,m_o = 1.80 \times 10^{-31}\,kg$), $B = 2.49 \times 10^2\,m^3\,s^{-2}$. The relative change of the velocity along the path could be expressed in function of the scaled distance $x = r/l_{Coul}$ as:

$$\frac{v(x)}{v_{th}} = \sqrt{1 + \left[\frac{exp(-\gamma x)}{x}\right]} \tag{6a}$$

where the scaling parameter is the Coulomb length, defined as:

$$l_{Coul} = \frac{B}{A} = \frac{e^2}{k_BT\,2\pi\varepsilon\varepsilon_o} \tag{6b}$$

i.e. the distance at which the thermal and Coulomb potential energies are equal. The Coulomb length $l_{Coul}$ is important control parameter describing the influence of Coulomb attraction on the capture time It is worth noting that $l_{Coul}$ is temperature dependent.

In order to grasp the value of this length the GaN values are given:

(i) for $T = 5K$ $\qquad\qquad l_{Coul} = 3.26 \times 10^{-7} m,$

(ii) for $T = 300K$ $\quad\quad\quad l_{Coul} = 5.42 \times 10^{-9} m$,

The larger value of $l_{Coul}$ denotes higher influence of the Coulomb force which is due to temperature decrease of thermal energy. Thus the acceleration is more important at low temperatures. Note also that the Coulomb length is much smaller than the capture length at room temperatures. At helium temperatures ($T \approx 5\ K$) for high point defect density, the capture and Coulomb lengths are comparable.

We introduce the parameter $\gamma$ as the screening factor that determines the degree of the screening at the Coulomb length. It is equal to the Coulomb length $l_{Coul}$ to screening distance $\lambda$ ratio:

$$\gamma = \frac{l_{Coul}}{\lambda} \tag{6c}$$

The parameter $\gamma$ controls the possible influence of screening of the Coulomb potential on the capture time:

i/ $\gamma \ll 1$

The screening factor $\gamma$ is close to zero, i.e. the screening length is much longer than Coulomb length, i.e. $\lambda \gg l_{Coul}$, therefore the screening is not affecting attraction potential, the acceleration is the same as for pure Coulomb potential.

ii/ $\gamma \gg 1$

In case the screening factor $\gamma$ is higher than unity, screening is very effective, therefore the capture time is determined by the thermal velocity alone, i.e. standard SRH approach could be used. It is worth to note that if the influence of the screening is very strong, it could overcome the acceleration by Coulomb force completely.

The magnitude of the screening distance can be determined from Debye screening length estimate [23, 24]:

$$\lambda = \sqrt{\frac{\varepsilon_o \varepsilon\ kT}{2e^2 n}} \tag{7}$$

where $n$ is the density of majority carriers. For the typically encountered densities these values are (for GaN at $T = 300K$):

i/ $n = 10^{17}\ cm^{-3}$ $\quad\quad \lambda = 1.21 \times 10^{-8}\ m$

ii/ $n = 10^{19}\ cm^{-3}$ $\quad\quad \lambda = 1.21 \times 10^{-9}\ m$

i/ $n = 10^{21}\ cm^{-3}$ $\quad\quad \lambda = 1.21 \times 10^{-10}\ m$

The first case corresponds to weakly doped semiconductor, the second - for heavily doped case while the latter - to highly excited active zone in light emitting diodes (LEDs) and laser diodes (LDs) [25, 26, 27, 28, 29, 30]. Because screening lengths are generally shorter than previously listed capture and Coulomb lengths, therefore the case corresponding to $\gamma \gg 1$ is most frequently encountered.

These two factors create a complex influence on the carrier velocity along the path, therefore the relative increase of the velocity in function of the scaled distance for different types of the screened potential is plotted in Fig. 1.

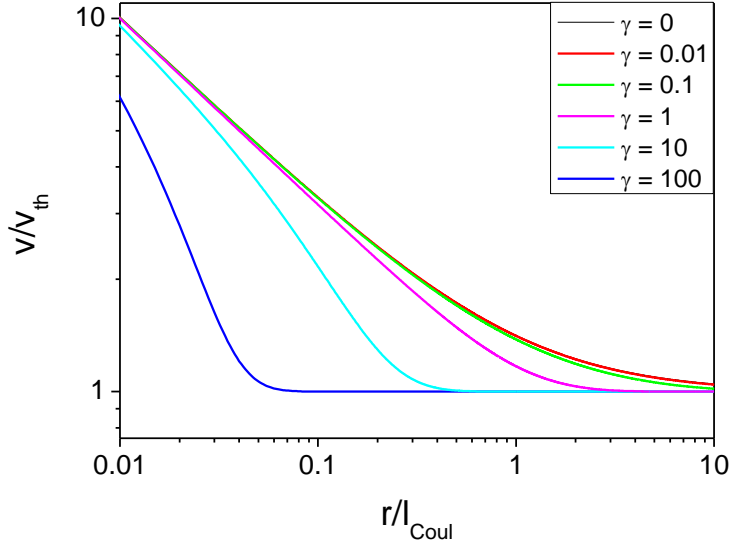

Fig. 1. Relative increase of the velocity of the carriers in the function of the scaled distance *x* from charged defect, calculated for several values of screening fraction $\gamma$. The value $\gamma = 0$ corresponds to Coulomb potential.

The data showed in Fig. 1 indicate that the velocity may be accelerated in drastically different manner. Nevertheless several general features can be distinguished. First, the noticeable acceleration emerges at the distances not higher than several Coulomb lengths. Second that the screening is extremely important, for the screening length close to Coulomb length the acceleration could be neglected at distances higher than the Coulomb length. Finally, at very close distances, the acceleration may increase carrier velocity by one order of magnitude.

These data indicate that the acceleration is important relatively close to the charged center and in the absence of effective screening. It is expected however that the carrier may travel longer distances. This leads to reduction of the capture time which cannot be estimated from the magnitude of the velocity directly. It has to be also assumed that the carrier has thermal velocity at the beginning of the path, i.e. for $r = l$. Thus the velocity is changed to:

$$\tilde{v}(r) = \sqrt{A + B\left[\frac{exp(-r/\lambda)}{r} - \frac{exp(-l/\lambda)}{l}\right]} \qquad (8)$$

Therefore the screened Coulomb attraction dependent capture time $\tau_{S-Coul}$ has to be calculated as the time of arrival of the carrier from the distance $l$ to the recombination center:

$$\tau_{S-Coul} = \int_0^l \frac{dr}{\tilde{v}(r)} = \int_0^l \frac{dr}{\sqrt{A+B\left[\frac{\exp(-r/\lambda)}{r} - \frac{\exp(-l/\lambda)}{l}\right]}}. \tag{9}$$

Since $l_{Coul} = B/A$ and $\gamma = l_{Coul}/\lambda$, expression (9) may be simplified by substituting $y = r/l$, followed by $\vartheta = l/l_{Coul}$. Then, the capture time (9) may expressed using the SRH capture time $\tau_{SRH} = \vartheta \frac{B}{A^{3/2}}$ and capture length to Coulomb length ratio $\vartheta$ as:

$$\tau_{S-Coul}(\vartheta) = \tau_{SRH} \int_0^1 \frac{dy}{\sqrt{1 - \frac{\exp(-\gamma\vartheta)}{\vartheta} + \frac{\exp(-\gamma\vartheta y)}{y\vartheta}}} \tag{10}$$

where we neglected the size of the defect as it is small, compared with the capture length, and additionally, the carrier velocity at the capture is very high due to acceleration and the integral was extended to zero, as its contribution is small due to very high velocity in this region.

First we consider the acceleration caused by influence of Coulomb potential only ($\gamma = 0$) on the capture time, (i.e. without screening - $\tau_{Coul}$) (10):

$$\tau_{Coul} = \tau_{SRH} \int_0^1 \frac{dy}{\sqrt{1 + \frac{1}{\vartheta}\left(\frac{1}{y} - 1\right)}} \tag{11}$$

The integral could be evaluated analytically. For $y > 0$, and constants $a, b > 0$, we have $\int \frac{dy}{\sqrt{a + \frac{b}{y}}} = \frac{\sqrt{ay^2 + b}}{a} - \frac{b\, \text{asinh}\left(\sqrt{\frac{ay}{b}}\right)}{a\sqrt{a}}$, and if additionally $ay \leq b$ then $\int \frac{dy}{\sqrt{-a + \frac{b}{y}}} = \frac{y\sqrt{y}}{\sqrt{b-ay}} - \frac{b\sqrt{y}}{a\sqrt{b-ay}} + \frac{b\, \text{asin}\left(\sqrt{\frac{ay}{b}}\right)}{a\sqrt{a}}$. Substituting $a = 1 - \frac{1}{\vartheta}, b = \frac{1}{\vartheta}$ for $\vartheta > 1$ in the first integral, $a = \frac{1}{\vartheta} - 1, b = \frac{1}{\vartheta}$ for $0 < \vartheta < 1$ in the second integral satisfying $ay \leq b$ for $0 \leq y \leq 1$, and for $\vartheta = 1$ using $\int \frac{dy}{\sqrt{\frac{1}{y}}} = \frac{2y\sqrt{y}}{3}$, then integrating over $y \in [0,1]$, the expression (11) gives:

$$\tau_{Coul}(\vartheta) = \begin{cases} \tau_{SRH} \frac{\vartheta}{\vartheta - 1}\left[1 - \frac{1}{\sqrt{\vartheta}\sqrt{\vartheta - 1}} \text{asinh}(\sqrt{\vartheta - 1})\right] & \vartheta > 1 \\ \frac{2}{3}\tau_{SRH} & \vartheta = 1 \\ \tau_{SRH} \frac{\sqrt{\vartheta}\text{asin}(\sqrt{1-\vartheta}) - \vartheta\sqrt{1-\vartheta}}{(1-\vartheta)^{3/2}} & 0 < \vartheta < 1 \end{cases} \tag{12}$$

This dependence could be converted into:

$$\tau_{Coul}(\vartheta) = \begin{cases} \tau_{SRH}\dfrac{\vartheta}{\vartheta-1}\left[1-\dfrac{1}{\sqrt{\vartheta}\sqrt{\vartheta-1}}\ln(\sqrt{\vartheta-1}+\sqrt{\vartheta})\right] & \vartheta > 1 \\ \dfrac{2}{3}\tau_{SRH} & \vartheta = 1 \\ \tau_{SRH}\dfrac{\sqrt{\vartheta}\,asin(\sqrt{1-\vartheta})-\vartheta\sqrt{1-\vartheta}}{(1-\vartheta)^{3/2}} & 0 < \vartheta < 1 \end{cases}$$

(13)

This result can be approximated as a function of the capture and Coulomb lengths ratio $\vartheta$ for two opposite regimes:

i/ weak attraction $l \gg l_{Coul}$ ($\vartheta \gg 1$)

$$\tau_{Coul} \cong \tau_{SRH}\left\{1-\dfrac{\ln(4\vartheta)}{2\vartheta}\right\} \tag{14a}$$

This relation recovers SRH capture time for very weak Coulomb attraction ( $l_{Coul}/l \to 0$).

ii/ strong attraction $l_{Coul} \gg l$ ($\vartheta \ll 1$)

$$\tau_{Coul} \cong \dfrac{\pi\tau_{SRH}}{2}\sqrt{\vartheta} \tag{14b}$$

This reflects reduction of the capture time for strong Coulomb attraction by the charged recombination center.

More complex picture emerges in the case of the screened Coulomb potential. The screened Coulomb capture time may be expressed as:

$$\tau_{S-Coul} = \tau_{SRH}\int_0^1 \dfrac{dy}{\sqrt{1-\dfrac{exp(-\gamma\vartheta)}{\vartheta}+\dfrac{exp(-\gamma\vartheta y)}{y\vartheta}}} \tag{15}$$

The integral in Eq. 15 cannot be evaluated analytically, therefore numerical integration was employed to obtain the capture time dependence on $\vartheta$ for various values of $\gamma$. The data obtained are plotted in Fig. 2.

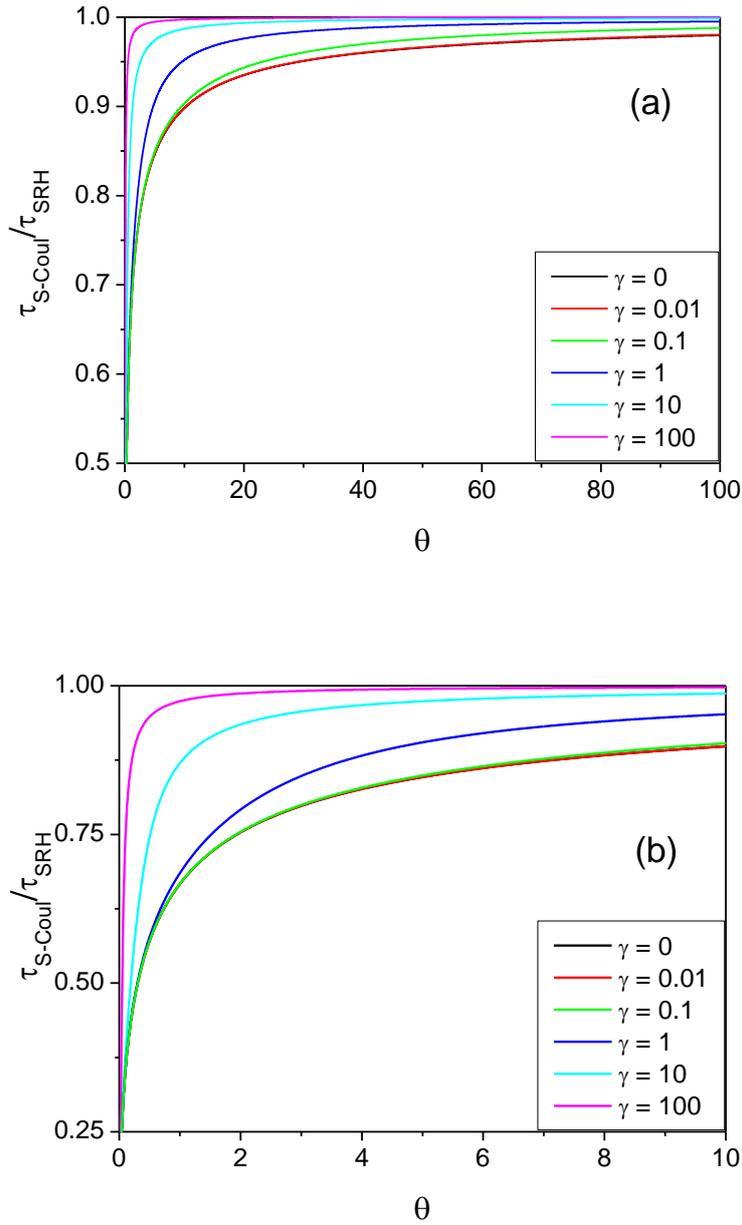

Fig. 2. Relative capture time (capture to SRH time ratio, i.e. $\tau_{S-Coul}/\tau_{SRH}$) in function of the scaled capture length (capture to Coulomb length ratio, i.e. $\vartheta = l/l_{Coul}$) for various values of screening factor $\gamma = \frac{l_{Coul}}{\lambda}$. $\gamma = 0$ corresponds to $\lambda \to \infty$, i.e. Coulomb attraction.

These results indicate that the Coulomb attraction affects the capture time over the large distances. The acceleration leads to shortening of the time even for the capture length as high $10^2$ Coulomb lengths. This is in agreement with the long range character of electrostatic interactions. Nevertheless the effect is not large, the capture time is shortened by 20% for 10

Coulomb lengths. On the other hand for close distance this reduction is more substantial: for capture length equal to Coulomb length the reduction is by one third. Drastic reduction is observed for very close distances, below $0.5\, l_{Coul}$. In contrast to long range behavior of Coulomb force, screening is governed by exponential law, i.e. it is short-ranged. This is reflected by the screened attraction data.

Wide and effective application of this formulation requires effective way of calculation of the capture time comparable to standard SRH estimate. In order to avoid somewhat cumbersome integration the approximate expression was obtained:

$$\tau_{S-Coul}(\vartheta) \cong \tau_{SRH} \left(\frac{\vartheta^{P1}}{\vartheta^{P1}+P2}\right)^{P3} \tag{16}$$

where the approximation parameters are given as functions of $\gamma$:

$$P1(\gamma) = 0.949 + 0.049\, log(\gamma) \tag{17a}$$

$$P2(\gamma) = 0.861 - 0.169\, log(\gamma) \tag{17b}$$

$$P3(\gamma) = 0.475 + 0.005327\, log(\gamma) + 0.001699\, [log(\gamma)]^2 \tag{17c}$$

These expression were verified to be used in the following range: $\gamma \in [0.01, 100.0]$. In total Eqs 16 and 17 recover accelerated time in the following range $\vartheta \in [0.01, 100]$ with the precision of 35 percent.

For ease of use of the provided expressions, both approximation (16) and formula (10), calculated by a numerical quadrature, are integrated into open-source Python library *unipress*. [31]

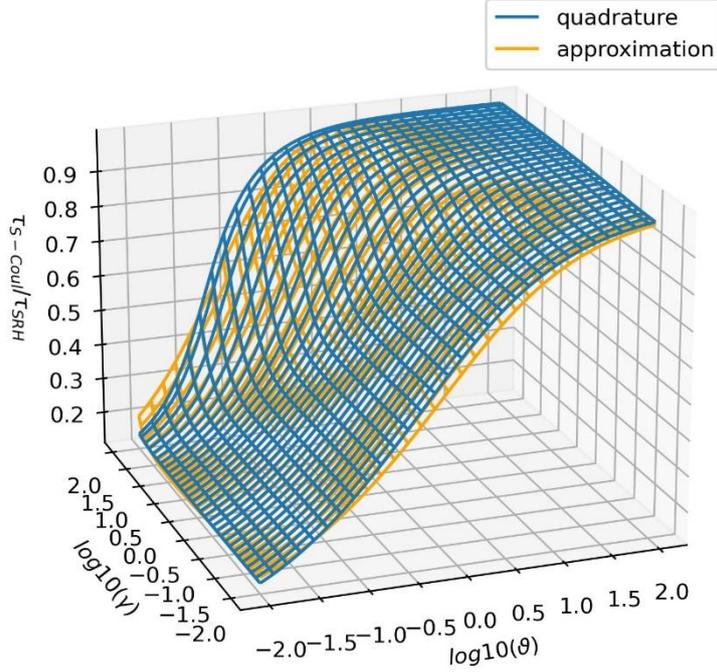

Fig. 3. Relative capture time ($\tau_{S-Coul}/\tau_{SRH}$) calculated by (10) by a numerical quadrature and by approximation (16).

The dependencies presented in Fig. 2 and the data quoted earlier could be used to assess the relative importance of the reduction of the capture time. As it is shown, in the pure Coulomb case without screening effects (i.e. $\gamma = 0$) the reduction is not limited to the Coulomb length i.e. to $\vartheta = l/l_{Coul} \approx 1$. For this case the reduction is of order of 40% (see Fig. 2b). Nevertheless it extends longer, for $\vartheta = l/l_{Coul} \approx 10$ this is still 20%, while in the case of $\vartheta = l/l_{Coul} \approx 100$ it is about 5% (see Fig. 2a). On the other hand, for relatively short capture length, such as $\vartheta = l/l_{Coul} \approx 0.1$ it is one order of magnitude smaller (see Fig. 3). Note that for $T = 5K$ we obtained $l_{Coul} = 3.26 \times 10^{-7} m$, therefore such extreme case can be achieved for high point defect densities, close to $N_{def} = 10^{20}\ cm^{-3}$. Thus the acceleration effects could be important for the low temperature PL investigations, where SRH recombination should not be neglected. Other important effect, i.e. screening could be also assessed form Fig. 2. As it is shown screening can reduce carrier acceleration especially for higher lengths. The typical behavior is for $\gamma = l_{Coul}/\lambda \approx 1$ in which the screening can reduce the Coulomb effect by the factor of 2. For shorter distances the screening is not as important. For higher $\gamma$ values, i.e. $\gamma = l_{Coul}/\lambda \approx 10$ or $\gamma = l_{Coul}/\lambda \approx 100$ the effect is very strong, reduces Coulomb effect by order of magnitude. Note that the screening length are shorter than the typical Coulomb or capture length, therefore such high factors are available especially for highly excited active layers in the devices. This effect, i.e. increase of SRH recombination rate for high excitation may explain LED efficiency droop observed for highly excited devices. [26, 32]

These effects should be investigated in more detail, which is planned in the further studies, including determination of SRH recombination rate in function of the point defect and dislocation densities, using newly developed method of analysis of time resolved photoluminescence (TRPL) signal decay described in Ref [13]. The planned research will include the role of the defects in the radiative recombination rates.

## IV. SUMMARY

It is shown that electrostatic attraction of the minority carriers by charged deep defect may reduce carrier capture time, significantly increasing SRH recombination rate. The effect is related to the rate limiting step, i.e. carrier capture in linear nonradiative (i.e. SRH) recombination. The reduction of the capture time is expressed in function of the two control parameters: capture length to Coulomb length ratio and the screening length to Coulomb length ratio. The Coulomb length is defined as the length at which the Coulomb energy is equal to the kinetic motion thermal energy, i.e. it is temperature dependent.

It is demonstrated that in the typical semiconductor device or semiconductor medium, the SRH recombination cannot be neglected at low temperatures. Effective screening by mobile carrier density could reduce the acceleration effect, leading to SRH decrease. The effect may be responsible for the decrease of LED efficiency at high excitation known as "droop effect".

The described effect of excited carriers' acceleration due to electrostatic attraction of the minority carriers by charged deep defect could enhance SRH recombination rates above the thermally imposed limits showing the possibility of the nonradiative recombination increase which were not explained by standard SRH formulation.

**Acknowledgements**

The research was partially supported by [National Science Centre, Poland]. This work was supported in part by the Collaborative Research Program of the Research Institute for Applied Mechanics, Kyushu University. Theoretical calculations were carried out with the support of the [Interdisciplinary Centre for Mathematical and Computational Modelling (ICM) University of Warsaw] under computational allocations no [G83-19] [GB76-25] and [GB84-23]. We gratefully acknowledge Poland's high-performance computing infrastructure PLGrid (HPC Centers: ACK Cyfronet AGH) for providing computer facilities and support within computational grant no. PLG/2022/015976.

# References

[1] W. Shockley and W. T. Read, "Statistics of the Recombinations of Holes and Electrons," *Physical Review,* vol. 87, p. 835–842, September 1952.

[2] S. Cuesta, A. Harikumar i E. Monroy, „Electron beam pumped light emitting devices," *Journal of Physics D: Applied Physics,* tom 55, p. 273003, July 2022.

[3] R. N. Hall, "Electron-Hole Recombination in Germanium," *Physical Review,* vol. 87, p. 387–387, July 1952.

[4] A. M. Stoneham, „Non-radiative transitions in semiconductors," *Reports on Progress in Physics,* tom 44, p. 1251–1295, December 1981.

[5] S. M. Sze and K. K. Ng, Physics of Semiconductor Devices, 1 ed., Wiley, 2006.

[6] C. Hamaguchi, Basic semiconductor physics, 2nd ed red., Berlin: Springer-Verlag, 2010.

[7] S. Chichibu, T. Azuhata, T. Sota and S. Nakamura, "Spontaneous emission of localized excitons in InGaN single and multiquantum well structures," *Applied Physics Letters,* vol. 69, p. 4188–4190, December 1996.

[8] S. F. Chichibu, A. C. Abare, M. S. Minsky, S. Keller, S. B. Fleischer, J. E. Bowers, E. Hu, U. K. Mishra, L. A. Coldren, S. P. DenBaars and T. Sota, "Effective band gap inhomogeneity and piezoelectric field in InGaN/GaN multiquantum well structures," *Applied Physics Letters,* vol. 73, p. 2006–2008, October 1998.

[9] K. P. O'Donnell, R. W. Martin and P. G. Middleton, "Origin of Luminescence from InGaN Diodes," *Physical Review Letters,* vol. 82, p. 237–240, January 1999.

[10] H. Murotani, Y. Yamada, T. Tabata, Y. Honda, M. Yamaguchi and H. Amano, "Effects of exciton localization on internal quantum efficiency of InGaN nanowires," *Journal of Applied Physics,* vol. 114, p. 153506, October 2013.

[11] H. Hirayama, N. Maeda, S. Fujikawa, S. Toyoda and N. Kamata, "Recent progress and future prospects of AlGaN-based high-efficiency deep-ultraviolet light-emitting diodes," *Japanese Journal of Applied Physics,* vol. 53, p. 100209, September 2014.

[12] V. K. Khanna, "Physical understanding and technological control of carrier lifetimes in semiconductor materials and devices: A critique of conceptual development, state of the art and applications," *Progress in Quantum Electronics,* vol. 29, p. 59–163, January 2005.

[13] P. Strak, K. Koronski, K. Sakowski, K. Sobczak, J. Borysiuk, K. P. Korona, P. A. Dróżdż, E. Grzanka, M. Sarzynski, A. Suchocki, E. Monroy, S. Krukowski and A. Kaminska, "Instantaneous decay rate

analysis of time resolved photoluminescence (TRPL): Application to nitrides and nitride structures," *Journal of Alloys and Compounds,* vol. 823, p. 153791, May 2020.

[14] M. Leszczynski, H. Teisseyre, T. Suski, I. Grzegory, M. Bockowski, J. Jun, S. Porowski, K. Pakula, J. M. Baranowski, C. T. Foxon and T. S. Cheng, "Lattice parameters of gallium nitride," *Applied Physics Letters,* vol. 69, p. 73–75, July 1996.

[15] S. Nakamura and S. F. Chichibu, Eds., Introduction to Nitride Semiconductor Blue Lasers and Light Emitting Diodes, 0 ed., CRC Press, 2000.

[16] T. Langer, H. Jönen, A. Kruse, H. Bremers, U. Rossow and A. Hangleiter, "Strain-induced defects as nonradiative recombination centers in green-emitting GaInN/GaN quantum well structures," *Applied Physics Letters,* vol. 103, p. 022108, July 2013.

[17] Ž. Gačević, A. Das, J. Teubert, Y. Kotsar, P. K. Kandaswamy, T. Kehagias, T. Koukoula, P. Komninou and E. Monroy, "Internal quantum efficiency of III-nitride quantum dot superlattices grown by plasma-assisted molecular-beam epitaxy," *Journal of Applied Physics,* vol. 109, p. 103501, May 2011.

[18] S. Watanabe, N. Yamada, M. Nagashima, Y. Ueki, C. Sasaki, Y. Yamada, T. Taguchi, K. Tadatomo, H. Okagawa and H. Kudo, "Internal quantum efficiency of highly-efficient $InxGa1-xN$-based near-ultraviolet light-emitting diodes," *Applied Physics Letters,* vol. 83, p. 4906–4908, December 2003.

[19] J. W. Orton i C. T. Foxon, „Group III nitride semiconductors for short wavelength light-emitting devices," *Reports on Progress in Physics,* tom 61, p. 1–75, January 1998.

[20] G. Martin, A. Botchkarev, A. Rockett and H. Morkoç, "Valence-band discontinuities of wurtzite GaN, AlN, and InN heterojunctions measured by x-ray photoemission spectroscopy," *Applied Physics Letters,* vol. 68, p. 2541–2543, April 1996.

[21] I. Vurgaftman, J. R. Meyer and L. R. Ram-Mohan, "Band parameters for III–V compound semiconductors and their alloys," *Journal of Applied Physics,* vol. 89, p. 5815–5875, June 2001.

[22] I. Vurgaftman and J. R. Meyer, "Band parameters for nitrogen-containing semiconductors," *Journal of Applied Physics,* vol. 94, p. 3675–3696, September 2003.

[23] W. Shockley, "The Theory of - Junctions in Semiconductors and - Junction Transistors," *Bell System Technical Journal,* vol. 28, p. 435–489, July 1949.

[24] W. Mönch, Semiconductor Surfaces and Interfaces, tom 26, G. Ertl, R. Gomer, H. Lüth i D. L. Mills, Redaktorzy, Berlin, Heidelberg: Springer Berlin Heidelberg, 2001.

[25] H. Yoshida, M. Kuwabara, Y. Yamashita, K. Uchiyama and H. Kan, "Radiative and nonradiative recombination in an ultraviolet GaN/AlGaN multiple-quantum-well laser diode," *Applied Physics Letters,* vol. 96, p. 211122, May 2010.

[26] X. Meng, L. Wang, Z. Hao, Y. Luo, C. Sun, Y. Han, B. Xiong, J. Wang and H. Li, "Study on efficiency droop in InGaN/GaN light-emitting diodes based on differential carrier lifetime analysis," *Applied Physics Letters,* vol. 108, p. 013501, January 2016.

[27] J. Piprek, "Efficiency droop in nitride-based light-emitting diodes," *physica status solidi (a),* vol. 207, p. 2217–2225, October 2010.

[28] A. David and M. J. Grundmann, "Droop in InGaN light-emitting diodes: A differential carrier lifetime analysis," *Applied Physics Letters,* vol. 96, p. 103504, March 2010.

[29] Q. Dai, M. F. Schubert, M. H. Kim, J. K. Kim, E. F. Schubert, D. D. Koleske, M. H. Crawford, S. R. Lee, A. J. Fischer, G. Thaler and M. A. Banas, "Internal quantum efficiency and nonradiative recombination coefficient of GaInN/GaN multiple quantum wells with different dislocation densities," *Applied Physics Letters,* vol. 94, p. 111109, March 2009.

[30] Y. C. Shen, G. O. Mueller, S. Watanabe, N. F. Gardner, A. Munkholm and M. R. Krames, "Auger recombination in InGaN measured by photoluminescence," *Applied Physics Letters,* vol. 91, p. 141101, October 2007.

[31] „unipress," [Online]. Available: https://github.com/ghkonrad/unipress.

[32] G. Verzellesi, D. Saguatti, M. Meneghini, F. Bertazzi, M. Goano, G. Meneghesso and E. Zanoni, "Efficiency droop in InGaN/GaN blue light-emitting diodes: Physical mechanisms and remedies," *Journal of Applied Physics,* vol. 114, p. 071101, August 2013.